\documentclass[review]{elsarticle}

\usepackage{lineno,hyperref}
\let\linenumbers\nolinenumbers\nolinenumbers

\usepackage[a4paper, total={6in, 8in}]{geometry}

\usepackage{multirow} 
\usepackage{makecell}
\usepackage{booktabs}
\usepackage{caption}

\usepackage{color}

\journal{arXiv}

\begin{document}

\begin{frontmatter}

\title{Domain generalization in deep learning for contrast-enhanced imaging
\tnoteref{mytitlenote}}

\author[1]{Carla Sendra-Balcells\corref{cor1}\fnref{fn1}}
\cortext[cor1]{Corresponding author:}
\ead{carla.sendra@ub.edu}
\fntext[fn1]{Shared first authorship.}

\author[1]{Víctor M. Campello\fnref{fn1}}
\author[1]{Carlos Martín-Isla}
\author[2]{David Vilades Medel}
\author[2]{Martín Luís Descalzo}
\author[3]{Andrea Guala}
\author[3]{José F. Rodríguez Palomares}
\author[1]{Karim Lekadir}

\address[1]{Dept. de Matemàtiques i Informàtica, Universitat de Barcelona, Spain}
\address[2]{Hospital de la Santa Creu i Sant Pau, Universitat Autònoma de Barcelona, Spain}
\address[3]{Cardiovascular Imaging Unit,  Hospital Universitari Vall d'Hebron, Barcelona, Spain}

\begin{abstract}
\textbf{Background:} 
The domain generalization problem has been widely investigated in deep learning
for non-contrast imaging over the last years, but it received limited attention
for contrast-enhanced imaging. However, there are marked differences in contrast
imaging protocols across clinical centers, in particular in the time between
contrast injection and image acquisition, while access to multi-center
contrast-enhanced image data is limited compared to available datasets for
non-contrast imaging. This calls for new tools for generalizing single-domain,
single-center deep learning models across new unseen domains and clinical
centers in contrast-enhanced imaging.\\
\textbf{Methods:} 
In this paper, we present an exhaustive evaluation of deep learning techniques
to achieve generalizability to unseen clinical centers for contrast-enhanced
image segmentation. To this end, several techniques are investigated, optimized
and systematically evaluated, including data augmentation, domain mixing,
transfer learning and domain adaptation. To demonstrate the potential of domain
generalization for contrast-enhanced imaging, the methods are evaluated for
ventricular segmentation in contrast-enhanced cardiac magnetic resonance imaging
(MRI).\\
\textbf{Results:} 
The results are obtained based on a multi-center cardiac contrast-enhanced MRI
dataset acquired in four hospitals located in three countries (France, Spain and
China). They show that the combination of data augmentation and transfer
learning  can  lead  to  single-center  models that  generalize  well  to  new
clinical  centers  not included during training.\\
\textbf{Conclusions:} 
Single-domain neural networks enriched with suitable generalization procedures
can reach and even surpass the performance of multi-center, multi-vendor models
in contrast-enhanced imaging, hence eliminating the need for comprehensive
multi-center datasets to train generalizable models.
\end{abstract}


\end{frontmatter}

\linenumbers

\section{Introduction}
\subsection{Problem and motivation}
Over the last years, the domain shift problem has attracted increased attention
in the medical image analysis community \cite{guan2021domain}. Several studies
have evaluated the level of generalization of deep learning techniques across
domains \cite{chen2020deep}. For example, a recent challenge on this topic was
organized in the cardiac magnetic resonance imaging (MRI) domain at the 2020
Medical Image Computing \& Computer-Assisted Intervention conference (MICCAI
2020), in collaboration with six Spanish, German and Canadian clinical centers.
Entitled "Multi-Centre, Multi-Vendor and Multi-Disease Cardiac Segmentation
(M\&Ms)", the study demonstrated that single-center, single-vendor neural
networks do not generalize naturally when segmenting cine-MRI images with
distinct imaging domains \cite{campello2021multi}. The lack of generalizability
of neural networks to unseen domains limits their clinical applicability at
scale. Thus far, several approaches have been attempted to address this problem
in non-contrast imaging, such as methods based on extensive spatial- and
intensity-based data augmentation \cite{chen2020improving}, the use of synthetic
images from generative models \cite{Kong2020AGD}, explicit domain adaptation (by
forcing the model to learn a similar representation across domains)
\cite{Parreno2021,Corral2021,scannell2020domain}, transfer learning
\cite{cheplygina2017transfer,kushibar2019supervised} and meta-learning
\cite{Liu2021SemisupervisedMW,Li2022DomainGO}. However, it is unclear whether
such approaches can improve generalizability in the case of complex imaging
modalities, such as in contrast-enhanced imaging, which is the subject of this
paper.

\begin{figure}[hbt!]
\centering
\includegraphics[width=14.cm]{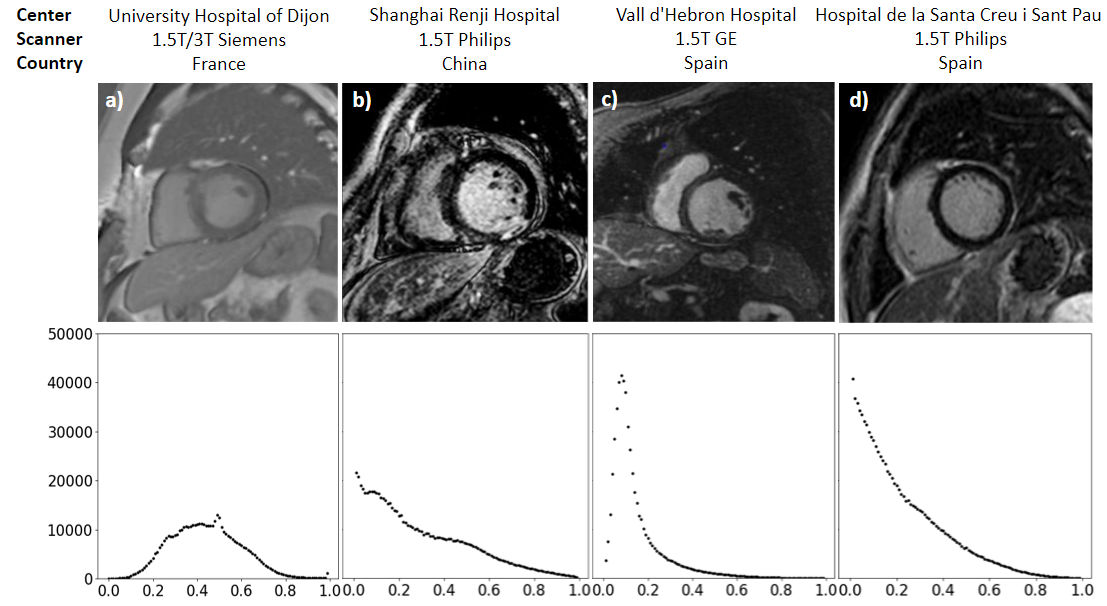}
\caption{\label{fig:mmm}Four LGE-MRI cardiac images acquired in four different
hospitals, together with the average intensity distribution of each dataset.
Each histogram has a very different shape and shows marked variability between
centers in terms of intensity distributions.}
\end{figure}

In many clinical applications, contrast-enhanced imaging is applied to further
improve the visibility of internal body structures and lesions in MRI
\cite{carr1984gadolinium}, Computed Tomography \cite{PEPE2020101773} or
Ultrasound \cite{otto2012practice} imaging. For example, late gadolinium
enhancement MRI (LGE-MRI) is an essential imaging modality for several
applications such as angiography \cite{riederer2018technical}, neuroimaging
\cite{ferre2012advanced}, oncology \cite{onishi2020ultrafast}, hepatology
\cite{welle2020mri} and cardiology \cite{uhlig2020gadolinium}. Contrast-enhanced
imaging is faced with additional challenges, compared to non-contrast imaging,
due to the intensity heterogeneity arising from the accumulation of the contrast
agent in the target areas and the artifacts introduced, which reduce the quality
of the images and modify the data distributions. Furthermore, the time between
contrast injection and image acquisition can vary greatly between patients and
centers, typically between 7 minutes up-to to a total of 10 minutes, resulting
in differences in contrast wash-out and image formation. As a result, the final
image appearance, both globally and locally, can have marked differences as
clearly illustrated in Figure \ref{fig:mmm} based on images from four different
hospitals. At the same time, the limited numbers of available LGE-MRI datasets
in existing open-access cohorts compared to non-contrast MRI images, combined
with legal and organizational obstacles across centers and countries, has made
access to interoperable multi-center LGE-MRI datasets more difficult. Hence,
there is a need for new tools for generalizing single-domain, single-center deep
learning models across new unseen domains and clinical centers in
contrast-enhanced imaging such as in LGE-MRI.

\subsection{Goals and contributions}
In this paper, we present an exhaustive evaluation of deep learning techniques
to achieve generalizability to unseen clinical centers for contrast-enhanced
imaging. To this end, several techniques are investigated, optimized and
systematically evaluated, including data augmentation, domain mixing, transfer
learning and domain adaptation. To demonstrate the potential of domain
generalization for contrast-enhanced imaging, the methods are evaluated for
ventricular segmentation in cardiac LGE-MRI \cite{doltra2013emerging}. For this
important clinical application, existing deep learning techniques have been
almost systematically trained and validated with an LGE-MRI sample from a single
clinical center
(\cite{yue2019cardiac,zabihollahy2020fully,kurzendorfer2019left,zhuang2020cardiac}).
As a result, while many research and commercial tools are already in use for
non-contrast cardiac MRI, image segmentation in cardiac LGE-MRI still relies on
labor-intensive manual delineation in clinical practice. Our work is based on a
unique multi-center cardiac LGE-MRI dataset acquired with three distinct scanner
vendors (Siemens, Philips and General Electric) in four hospitals located in
three countries (France, Spain and China).

\section{Methods}
\begin{figure}[hbt!]
\centering
\includegraphics[width=14.3cm]{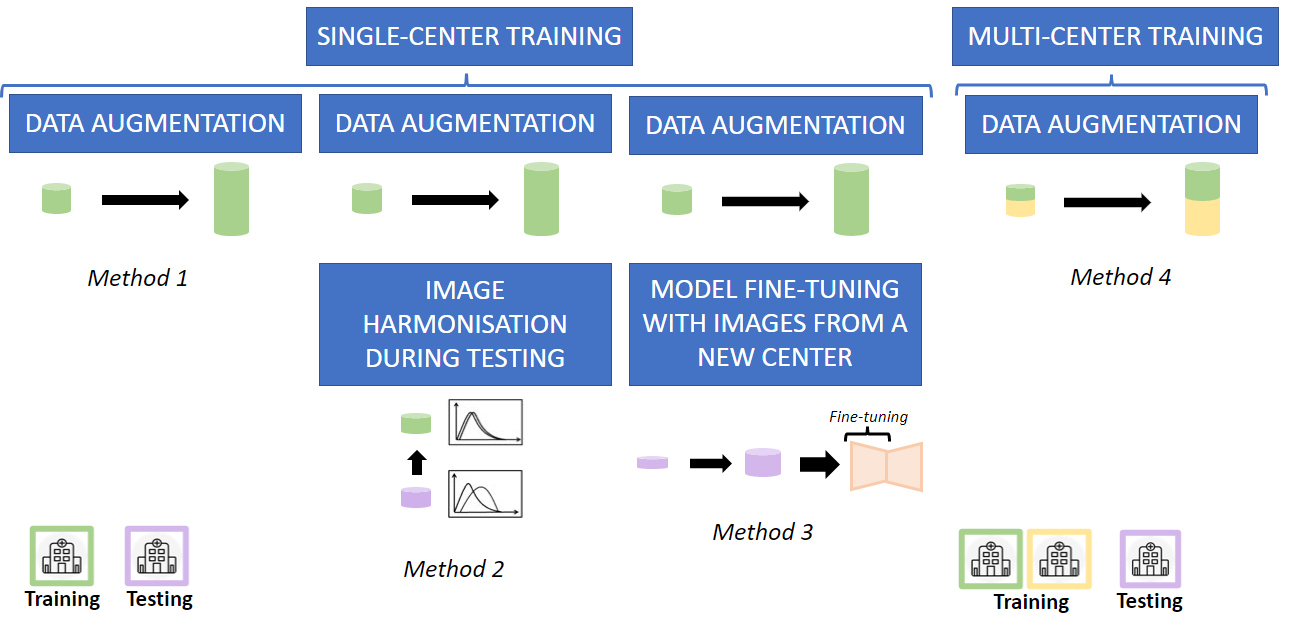}
\caption{\label{fig:generalization techniques}Four different approaches
implemented in this work to enhance the generalisability of LGE-MRI segmentation
models across distinct clinical sites.}
\end{figure}

In this section, an end-to-end pipeline is investigated for generalizable image
segmentation in multi-center  LGE-MRI datasets. It is applied for deep
learning-based segmentation of the left ventricle (LV), including the blood pool
and the myocardium, in multi-center LGE-MRI  cardiac images. To this end, four
different approaches are explored to enhance the generalizability across
clinical sites of existing deep neural networks for LGE-MRI segmentation, as
schematically represented in Figure \ref{fig:generalization techniques}. These
include: 

\begin{enumerate}
\item Data augmentation techniques to artificially extend the data distribution
captured by the trained models.
\item Image harmonization to align the data distributions of the training and
testing images.
\item Transfer learning to adjust the neural network to the new clinical center
based on very few unseen images.
\item Multi-center models directly trained with data from multiple clinical
centers, which are used for comparative evaluation of the different
generalization mechanisms.
\end{enumerate}

We confirm that all experiments were performed in accordance with relevant
guidelines and regulations.

\subsection{Datasets}
The multi-center and multi-vendor dataset used in this study consists of 216
cardiac LGE-MRI datasets acquired in four different clinical centers as detailed
in Table \ref{tab:dataset summary}. Two out of four samples are publicly
available datasets from France and China, while the two other samples correspond
to new LGE-MRI images acquired in two different hospitals in Spain. The subjects
have been scanned by using a range of scanner vendors by Siemens, Philips or
General Electric (GE). In addition to having distinct intensity distributions as
observed in Figure \ref{fig:mmm}, the multi-center LGE-MRI images also differ in
the image resolution (0.75-1.88 mm), slice thickness (5-13 mm), and acquisition
time after contrast injection (7 to 10 minutes). The samples from each clinical
site are described in more detail in the next subsections. 

\renewcommand{\arraystretch}{1.5}
\begin{table}
    \captionof{table}{Details of the multi-center LGE-MRI datasets and characteristics of the acquired images used in this work. Imaging time = Acquisition time after contrast injection.}
    \scriptsize
    \centering
    \begin{tabular}{|l|l|l|l|r|r|r|r|r|}
    \hline
    \textbf{Dataset} & \textbf{\makecell{Clinical\\ center}} & \textbf{Country}
    & \textbf{\makecell{MRI\\ scanner}} & \textbf{\makecell{Imaging\\ time\\
    (mins)}} & \textbf{\makecell{In-plane\\ resolution\\ (mm)}} &
    \textbf{\makecell{Slice\\ thickness\\ (mm)}} & \textbf{\makecell{Number\\ of
    slices}} & \textbf{\makecell{Sample\\ size}}\\ 
    \hline
    EMIDEC & \makecell{University\\ Hospital\\ of Dijon} & France &
    \makecell{1.5T\\ and 3T\\ Siemens} & 10 & 1.37-1.88 & 8-13 & 4-10 & 100\\
    \hline
    MSCMR & \makecell{Shanghai\\ Renji\\ Hospital} & China & \makecell{1.5T\\
    Philips} & - & 0.75 & 5 & 10-18 & 45\\ 
    \hline
    VH & \makecell{Vall\\ d\textquotesingle Hebron\\ Hospital} & Spain & 1.5T GE
    & 10 & 1.48-1.68 & 10 & 8-15 & 41\\
    \hline
    STPAU & \makecell{Sant Pau\\ Hospital} & Spain & \makecell{1.5T\\ Philips} &
    7-10 & 1.18 & 5 & 18-24 & 30\\
    \hline
    \end{tabular}
\label{tab:dataset summary}
\end{table}

\subsubsection{EMIDEC dataset: University Hospital Dijon, France}
This dataset was compiled as part of the automatic Evaluation of Myocardial
Infarction from Delayed-Enhancement Cardiac MRI challenge (EMIDEC)
\cite{lalande2020emidec}. The EMIDEC volunteers included  33 healthy and 67
diseased subjects, for a total of 100 studies. The data was acquired at the
University Hospital of Dijon, France, using Area 1.5 T as well as Skyra 3T
Siemens MRI scanners. Slice thickness and in-plane spatial resolution varied
greatly, being comprised between 8 and 13 mm and 1.37 and 1.88 mm, respectively.
The manual segmentation of the LV blood pool and myocardium were performed by a
cardiologist with over 10 years of experience. It is the largest of the four
samples and hence it was used as the reference sample for training the
single-center neural networks.

\subsubsection{MSCMR dataset: Shanghai Renji Hospital, China}

The MSCMR dataset was obtained from the Multi-sequence Cardiac MR Segmentation
Challenge and it comprises a total of 45 patients suffering from various
cardiomyopathies (\cite{zhuang2016multivariate}, \cite{zhuang2018multivariate}).
The images were acquired at the Shanghai Renji hospital, China, which will allow
us to evaluate generalizability across countries as well as continents in this
study. Compared to EMIDEC, the MSCMR dataset has a higher image resolution
(in-plane resolution = 0.75 mm, slice thickness = 5 mm for all scans) and all
images were acquired with a 1.5 T Philips scanner. The manual delineations were
initially performed by trainees and later on validated by expert cardiologists.

\subsubsection{VH dataset: Vall d'Hebron Hospital, Spain}

The VH dataset consists of 41 LGE-MRI datasets acquired at the Vall
d\textquotesingle Hebron University Hospital, located in Barcelona, Spain. In
addition to covering a new geographical location, namely Spain, the VH sample
has several differences with EMIDEC and MSCMR, including the disease group (MI)
and the MRI scanner (1.5 T GE scanner). Manual annotations of the LV boundaries
were generated by a trained rater using the cvi42 software. The study was
approved by the ethics committee of the Vall d'Hebron Hospital and written
informed consent was obtained from all participants.

\subsubsection{STPAU dataset: Sant Pau Hospital, Spain}
The STPAU dataset comprises 30 LGE-MRI cases acquired at the Sant Pau Hospital
in Barcelona, Spain. While the clinical center is located in the same region as
for the VH sample, the dataset covers a different disease group (ischemic and
non-ischemic cardiomyopathy) and was acquired using an MRI scanner from a
different vendor (Philips Achieva 1.5T) and a higher-resolution imaging
protocol. Furthermore, the time delay between contrast injection and image
acquisition varies between 7 and 10 minutes, which adds extra variability. The
manual annotations were also performed using cvi42, as in the previous case. All
patients signed the informed consent, the study protocol was approved by the
Ethical Committee for Clinical Research of our region, and it follows the
ethical guidelines of the Declaration of Helsinki.

\subsection{Single-center model with data augmentation}

\begin{figure}[hbt!]
\centering
\includegraphics[width=12cm]{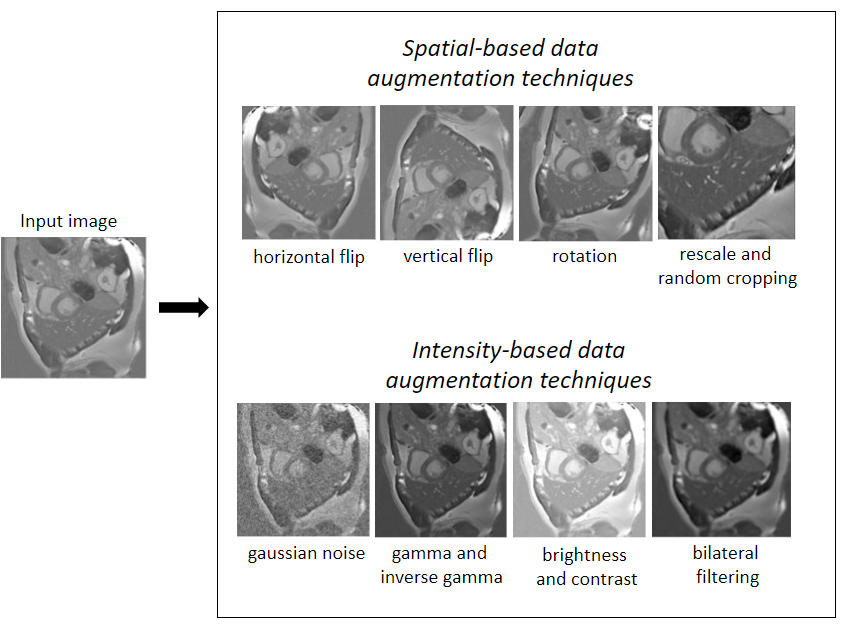}
\caption{\label{fig:data augmentation pipeline}Both spatial and intensity-based
data augmentation techniques are applied together with a probability of 0.2
each. From only one slice many samples can be generated, increasing the size of
the original dataset significantly.}
\end{figure}

In this work, we first investigate the potential of data augmentation to enhance
the generalizabilty of LGE-MRI segmentation models (Method 1 in Figure
\ref{fig:generalization techniques}). Data augmentation has been widely used to
create more robust neural networks by increasing the size as well as the
heterogeneity of training samples synthetically. However, the promise of data
augmentation is yet to be examined for LGE-MRI, where there is higher complexity
due to inherent variability in scar characteristics and contrast appearance. 

In this work, we investigate several operators for data augmentation in the
context of LGE-MRI as illustrated in Figure \ref{fig:data augmentation pipeline}
and described as follows.
\begin{enumerate}
    \item \textbf{Spatial-based data augmentation}: In addition to the natural
    variability between cardiac anatomies, especially across countries and
    ethnic groups, patients undergoing LGE-MRI typically suffer from regional
    remodelling of the ventricles due to the presence of scar tissue. Hence,
    spatial-based data augmentation is proposed using the following operators:
    \begin{itemize}
        \item Horizontal and vertical flips to generate images with different
        orientations.
        \item Random rotations of up to $\pm30$ degrees, to simulate different
        positions of the heart.
        \item Random rescaling in the [0.75, 1.88] mm range so that the model
        can process images and hearts that vary in size. This range is defined
        by the minimum and maximum voxel size of our multi-center dataset.
        \item Random cropping, such that the training images have the same
        dimensions of 256x256 pixels but with a variation in the position of the
        heart in the image.
    \end{itemize}
    
    \item \textbf{Intensity-based data augmentation:} Because the LGE-MRI
    appearance can vary between images acquired using different MRI scanners and
    scanning protocols, such as due to differences in acquisition time after
    contrast injection, we implemented the following intensity-based data
    augmentation techniques:
    
    \begin{itemize}
        \item Bilateral filtering to generate blurred and less detailed copies
        of the original images. 
        \item Gaussian noise with a standard deviation ranging between [0, 0.03]
        to generate artificial noise and image artifacts.
        \item Gamma and inverse Gamma function with magnitude [0.7, 1.5] to
        generate synthetic images with different lighting.
        \item Brightness and contrast with magnitude [-0.5, 0.5] to support
        brightness and contrast variations in the training images.
    \end{itemize}
    
\end{enumerate}

Each data augmentation technique is applied with a probability of 0.2 during the
training of the model. Then, this data augmentation pipeline is evaluated by
measuring the final generalization ability of the network (Method 1 in Figure
\ref{fig:generalization techniques}). Table \ref{tab:dataset data augmentation}
summarizes the split of the data used for the training, validation and testing
of the model in each experiment. 

\renewcommand{\arraystretch}{1.5}
\begin{table}
    \captionof{table}{Number of subjects for each of the four datasets used during the training, validation and testing phases when data augmentation is implemented in a single-center setting.}
    \scriptsize
    \centering
    \begin{tabular}{ |p{1.5cm}|p{1.5cm}|p{1.5cm}|p{1.5cm}|}
    \hline
    \textbf{Dataset} & \textbf{Training} & \textbf{Validation} &
    \textbf{Testing}\\ 
    \hline
    EMIDEC & 68 & 17 & 15\\
    \hline
    MSCMR & 24 & 6 & 15\\ 
    \hline
    VH & 21 & 5 & 15\\
    \hline
    STPAU & 12 & 3 & 15\\
    \hline
    \end{tabular}
\label{tab:dataset data augmentation}
\end{table}

\subsection{Image harmonization at testing}
\begin{figure}[hbt!]
\centering
\includegraphics[width=14.cm]{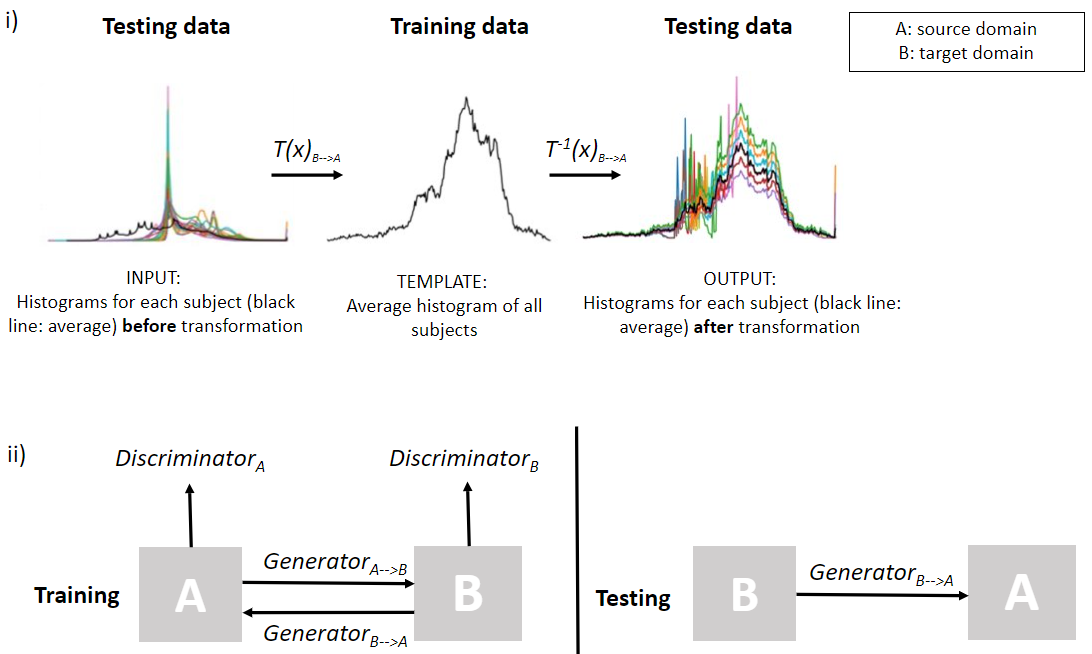}
\caption{\label{fig:data harmonization}Schematic illustration of the image
harmonisation techniques used in this work to make the intensity distributions
from the different clinical sites as aligned as possible. First, histogram
matching is used to learn a transformation of the histogram of each image from
the unseen clinical sites (B) onto the original training clinical center (A).
Second, CycleGAN architecture is used to learn the mapping between the training
and the testing clinical center.}
\end{figure}

While the data augmentation operations focused on improving model
generalizability at training, we propose to apply image harmonization at the
testing stage to further reduce the discrepancies between the multi-center
LGE-MRI images (Method 2 in Figure \ref{fig:generalization techniques}). Image
harmonization enables to transform the testing LGE-MRI images from a new
clinical center such that their intensity distribution matches as much as
possible the imaging characteristics of the single center used to train the
baseline neural network. Concretely, two main image harmonization techniques
were implemented:

\begin{enumerate}
    \item \textbf{Histogram matching}: It consists of transforming the testing
    images from the unseen center such that the histogram of the pixel intensity
    values is superimposed as much as possible with the corresponding histogram
    extracted from the training images from the training clinical center. The
    transformation from the testing data (B: target data) to the training data
    (A: source data) is illustrated in Figure \ref{fig:data harmonization}(i). 
    \item \textbf{CycleGAN}: Another strategy to address the domain shift
    between multiple centers is domain adaptation, which can be used to learn
    the image translation from the source domain to the target domain. To this
    end, we choose to implement a CycleGAN architecture \cite{zhu2017unpaired},
    based on an unpair image-to-image translation. Given that CycleGAN uses
    cycle consistency, it would learn the translation from the target domain (B)
    to the source domain (A), and viceversa (Figure \ref{fig:data
    harmonization}(ii)). Both target to source and source to target generators
    are saved in each implementation, decreasing to 6 the number of
    implementations needed. The amount of samples used to train each of the
    CycleGAN models are summarised in Table \ref{tab:dataset CycleGAN},
    adjusting for each case the percentage of images from each center so that it
    is adequately balanced (50\% source and 50\% target data).  
\end{enumerate}

\renewcommand{\arraystretch}{1.5}
\begin{table}
    \captionof{table}{Number of samples used for the training and validating each CycleGAN model built to harmonise the imaging properties from the different clinical centers.}
    \scriptsize
    \centering
    \begin{tabular}{|*{6}{c|}}
    \hline
    \multicolumn{2}{|c|}{\textbf{Dataset}} &
    \multicolumn{2}{c|}{\textbf{Training}} &
    \multicolumn{2}{c|}{\textbf{Validation}}\\
    \hline
    Source & Target & Source & Target & Source & Target\\
    \hline
    EMIDEC & MSCMR & 24 & 24 & 6 & 6\\
    \hline
    EMIDEC & VH & 21 & 21 & 5 & 5\\ 
    \hline
    EMIDEC & STPAU & 12 & 12 & 3 & 3\\
    \hline
    MSCMR & VH & 21 & 21 & 5 & 5\\
    \hline
    MSCMR & STPAU & 12 & 12 & 3 & 3\\
    \hline
    VH & STPAU & 12 & 12 & 3 & 3\\
    \hline
    \end{tabular}
\label{tab:dataset CycleGAN}
\end{table}

\subsection{Transfer learning from the original to the new clinical site}
Another strategy investigated in this work to improve the scalability of
single-center models consisted of applying the so-called transfer learning
paradigm, by fine-tuning specific layers of the neural network with a reduced
number of LGE-MRI images from the new clinical site (Method 3 in Figure
\ref{fig:generalization techniques}). The approach has shown promise for
multi-center image segmentation in cardiac cine-MRI \cite{ma2019neural}, but is
yet to be demonstrated for multi-center LGE-MRI imaging, where there is
increased variability. The following steps are implemented in this work:

\begin{enumerate}
    \item Initiate the training of the neural network with the EMIDEC dataset,
    then evaluate the minimum number of fine-tuned layers, in both the decoder
    and the encoder, that are needed during transfer learning to obtain the
    maximal segmentation performance on the new LGE-MRI datasets from the
    remaining clinical centers.  
    \item Compare the previous results with the segmentations obtained based on
    a multi-center model directly trained each time with images from two
    clinical centers (EMIDEC and the new center).
    \item Estimate the minimum percentage of images needed from the second
    clinical center during the fine-tuning to obtain the desired level of
    performance.
    \item Implement the same approach from the previous point but this time by
    using a model pre-trained on a large dataset (n=350) from cine-MRI (M\&Ms
    dataset), to evaluate transfer learning from a related cardiac MRI modality
    for which data is abundantly available. 
\end{enumerate}

\subsection{Multi-center model}

A fourth and last modelling strategy, i.e. training the neural networks directly
from multiple centers (Method 4 in Figure \ref{fig:generalization techniques}),
is used for comparative evaluation of the three extended single-center models
described in the previous section, i.e. enriched with data augmentation, image
harmonization and transfer learning. In this study, we investigated the amount
of new centers/domains that are needed to bridge the domain gap in LGE-MRI
segmentation, by using a balanced dataset with the same number of subjects for
each multi-center data combination, namely EMIDEC, EMIDEC+MSCMR, EMIDEC+VH,
EMIDEC+MSCMR+VH and ALL centers. The samples used for training the multi-center
models in each combination of datasets/centers are listed in Table
\ref{tab:dataset data harmonization}. In all experiments, the same testing
dataset is used for comparative evaluations (n=15). 

\renewcommand{\arraystretch}{1.5}
\begin{table}
    \captionof{table}{List and number of samples used for training and validating multi-center models in this study.}
    \scriptsize
    \centering
    \begin{tabular}{ |l|r|r|}
    \hline
    \textbf{Dataset} & \textbf{Training} & \textbf{Validation}\\
    \hline
    EMIDEC & 42 & 10\\
    \hline
    EMIDEC+MSCMR & 21+21 & 5+5\\
    \hline
    EMIDEC+VH & 21+21 & 5+5\\ 
    \hline
    EMIDEC+MSCMR+VH & 14+14+14 & 3+3+3\\
    \hline
    EMIDEC+MSCMR+VH+STPAU & 11+11+11+11 & 3+3+3+3\\
    \hline
    \end{tabular}
\label{tab:dataset data harmonization}
\end{table}

\subsection{Baseline workflow}

\subsubsection{Pre-processing}

Min-max normalization is used after cropping of the image to keep the same
intensity range in images from the same dataset. 

\subsubsection{Post-processing}

A post-processing is applied to all predictions generated by the model by
keeping only the largest connected component of the segmentation volume. This
step is commonly used in medical image segmentation, especially in organ
imaging, to help on the detection of false positives.  

\subsubsection{Network architecture}

\begin{figure}[hbt!]
\centering
\includegraphics[width=14cm]{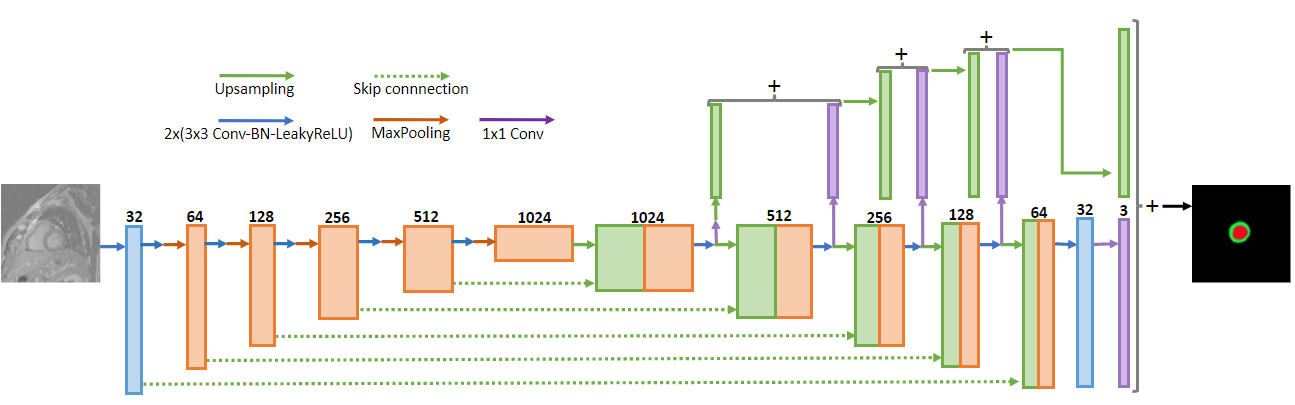}
\caption{\label{fig:network}U-Net architecture composed by 6 layers, increasing
progressively the number of feature maps until 1024. Additionally, deep
supervision layers are included in the decoder.}
\end{figure}

As a baseline model, a U-Net architecture is implemented to perform the LV
boundary segmentations in LGE-MRI based on some of the modifications proposed by
\cite{isensee2021nnu} for improved model training as follows. First, Leaky ReLU
is used as activation function, then instance normalization is applied after
each hidden convolutional layer to stabilise the training. Deep supervision is
included to allow gradients to be injected deeper into the network and
facilitating the training of all layers. Furthermore, a 2D architecture is
selected as it is suitable to address the differences in slice thickness between
clinical centers, as well as slice misalignment due to respiratory and cardiac
motion artefacts. The encoder and decoder architecture of the model are
illustrated in Figure \ref{fig:network}.

\subsubsection{Implementation details}

PyTorch is the open-source machine learning library for Python used for the
implementation of the model learning. Stochastic gradient descent (SGD)
optimization is performed with Adam and the batch size of 16 slices is
constrained by the 8 GB of memory of the NVIDIA GeForce RTX 2080 Ti GPU. The
learning rate is kept to 1$\cdot$10$^{-3}$ during every training, while the dice
and cross entropy losses are calculated at every iteration to optimize the
network parameters. The neural network is trained 250 epochs each time and takes
half an hour approximately to converge. During testing, each LGE-MRI image
segmentation takes less than one second. The main criterion followed to split
each dataset in subgroups is 80\% for the training and 20\% for the validation
part, while keeping 15 subjects for the testing.

\subsubsection{Performance evaluation}

For all experiments and results, the performance of each method will be assessed
with the average 3D Dice Coefficient (DC), which calculates the overlap ratio
between the automatically generated and ground truth segmentations. The measure
is estimated by:
\begin{equation}
    \mbox{DC} = \frac{2 \cdot (X \cap Y)}{X + Y} = \frac{2 \cdot TP}{2 \cdot TP + FP + FN},
\end{equation}
where X and Y are the set of pixels from the automated and true labels of the
target structures, while TP, FP, and FN are the corresponding true positives,
false positives and false negatives, respectively.

\section{Results}

This section presents detailed experimental results obtained by evaluating and
comparing the different strategies proposed for enhancing model generalizability
in LGE-MRI segmentation. Four experiments are proposed to study model
generalizability: (1) effect of data augmentation, (2) image harmonization, (3)
transfer learning and (4) multi-center training. The results are summarized in
Table~\ref{tab:results_summary}, where a similar limited generalization
performance is achieved for experiments (1) and (2) while experiments (3) and
(4) show a significant improvement. Each experiment is analyzed in detail next.

\begin{table}
    \scriptsize
    \centering
    \caption{Dice score coefficient for the different domain generalization
    experiments performed. The results are averaged over five runs of models.
    All models used EMIDEC for training. In experiment 3, every model is
    transferred to the corresponding target center and in experiment 4, every
    model is trained with EMIDEC and a training set from the corresponding
    target center. Standard deviation is presented as subscript for five
    independent runs of each model.}
    \begin{tabular}{lccccccc}
        \toprule
        {} & \multicolumn{3}{c}{\emph{Experiment 1}} & \multicolumn{2}{c}{\emph{Experiment 2}} & \emph{Experiment 3} & \emph{Experiment 4} \\
        Test Center & \multicolumn{3}{c}{\makecell{Effect of data augm.\\(single-center training)}} & \multicolumn{2}{c}{Image harmonization} & Tranfer learning & \makecell{Multi-center\\ training} \\
        {} &   No augm. & Spatial & \makecell{Spatial \&\\ intensity} & CycleGAN & Hist. match. \\
        \midrule
        EMIDEC & $0.85_{0.05}$ & $0.88_{0.03}$ & $0.78_{0.08}$ &  - &    - &  - &  - \\
        MSCMR  & $0.30_{0.15}$ & $0.62_{0.19}$ & $0.72_{0.12}$ & $0.64_{0.17}$ & $0.78_{0.07}$ & $0.87_{0.03}$ & $0.89_{0.03}$ \\
        STPAU  & $0.54_{0.16}$ & $0.61_{0.12}$ & $0.68_{0.09}$ & $0.70_{0.08}$ & $0.68_{0.08}$ & $0.85_{0.04}$ & $0.85_{0.04}$ \\
        VH     & $0.32_{0.21}$ & $0.26_{0.23}$ & $0.62_{0.13}$ & $0.53_{0.17}$ & $0.58_{0.12}$ & $0.78_{0.11}$ & $0.82_{0.06}$ \\
        \bottomrule
    \end{tabular}
\label{tab:results_summary}
\end{table}

\subsection{Experiment 1: Effect of data augmentation}
\begin{figure}[hbt!]
    \centering
    \includegraphics[width=.9\textwidth]{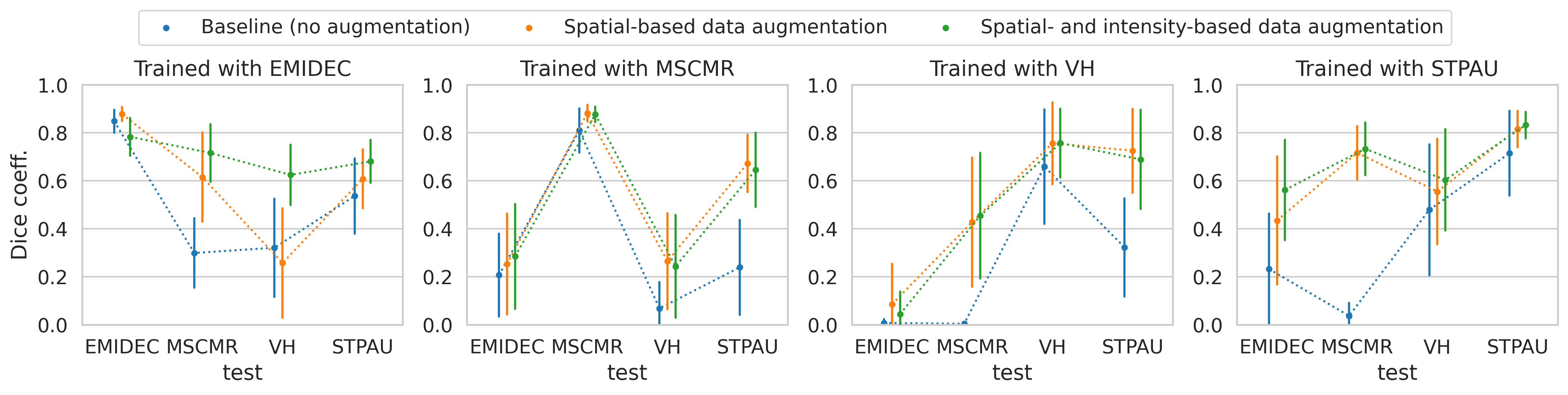}
    \caption{Comparison of the subject-wise Dice coefficient obtained for models
    trained with a single-center with and without data augmentation, including
    spatial- and intensity-based augmentations. Models are tested on subjects
    from the testing set for every center. The results are averaged over five
    different runs of each model.}
    \label{fig:generalization}
\end{figure}

In the first experiment, the added value of the different types of data
augmentation is evaluated, including spatial and intensity-based data
augmentations. Figure \ref{fig:generalization} shows the comparative results
obtained by three different models: (i) a single-center model without data
augmentation (blue line), (ii) a single-center model with spatial data
augmentation (orange), and (iii) a single-center model enriched with both
spatial and intensity based data augmentations (green). As observed in the
results, data augmentation consistently improves the segmentation performance
for LGE-MRI independently of the clinical center used for training, increasing
the DC value up to 0.6 units with respect to the baseline model without data
augmentation. Furthermore, the results in Figure \ref{fig:generalization} show
that while most of the improvement can be achieved by spatial data augmentation
(orange line), intensity-based data augmentation adds value to the approach, in
particular when training on the largest sample (EMIDEC) and testing on smaller
samples (MSCMR, VH and STPAU). Having demonstrated the added value of data
augmentation, all subsequent experiments are performed using spatial- and
intensity-based data augmentation.

\subsection{Experiment 2: Effect of image harmonization}

\begin{figure}[hbt!]
    \centering
    \includegraphics[width=.9\textwidth]{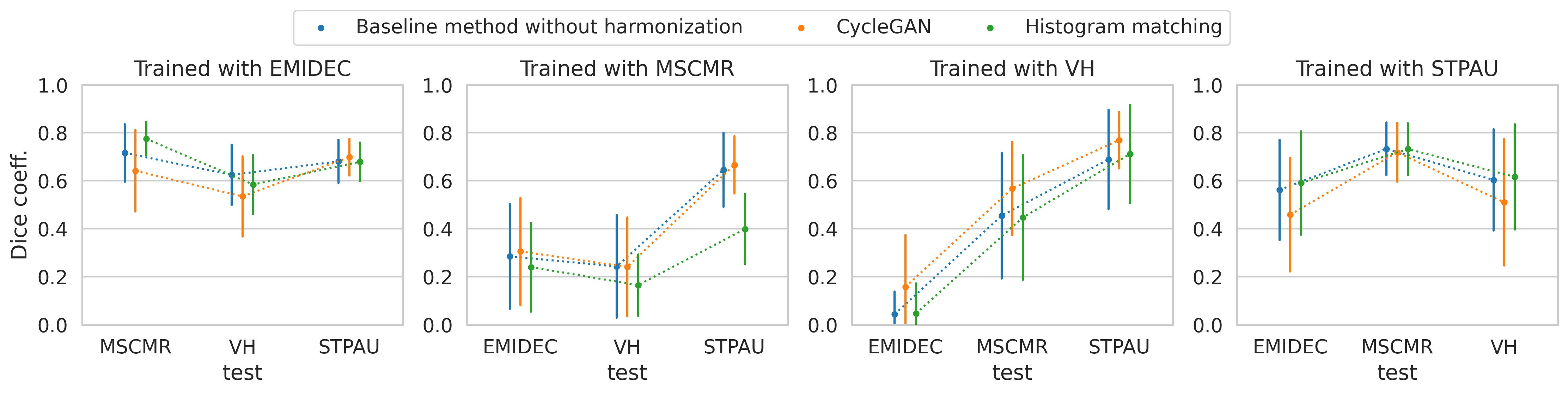}
    \caption{Effect of histogram matching and CycleGAN harmonisation for LGE-MRI
    segmentation in unseen clinical centers. X corresponds to each testing
    center not included in the training dataset and the results are averaged
    over five different runs of each model.}
    \label{fig:pipeline}
\end{figure}

Here, the impact of image harmonization is evaluated when applied to match the
intensity distribution and appearance of LGE-MRI images from a new clinical
center to that of the training set. Specifically, we evaluate three approaches,
namely (i) the baseline model with data augmentation from Experiment 1 but
without any normalization, (ii) the baseline model with histogram matching, and
(iii) the baseline model with CycleGAN normalization. The results are given in
Figure \ref{fig:pipeline}, clearly showing that, overall, the two harmonization
operations (green and orange lines) do not improve significantly the LGE-MRI
segmentations over the baseline model without harmonization (blue line). There
are, however, few cases where the mean Dice score is sligthly improved, as for
histogram matching when training with EMIDEC and testing in MSCMR or when using
CycleGAN for models trained with VH.

\subsection{Experiment 3: Effect of transfer learning}
\begin{figure}[hbt!]
    \centering
    \includegraphics[width=.9\textwidth]{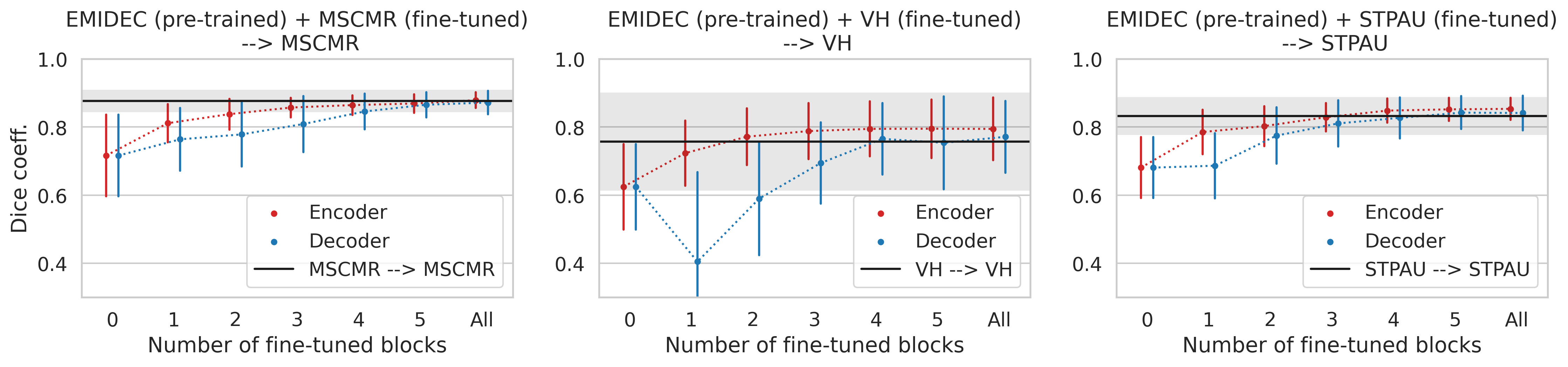}
    \caption{Evaluation of a single-center model pre-trained on the EMIDEC
    dataset and fine-tuned with a new clinical dataset (MSCMR, VH or STPAU).
    Red: Fine-tuning of a number of blocks in the encoder. Blue: Fine-tuning of
    a number of blocks in the decoder. Black: Model trained from scratch with
    data from the same center. Bars and gray band stand for the
    standard deviation of the five independent model runs.}
    \label{fig:blocks}
\end{figure}

\begin{figure}[hbt!]
    \centering
    \includegraphics[width=.8\textwidth]{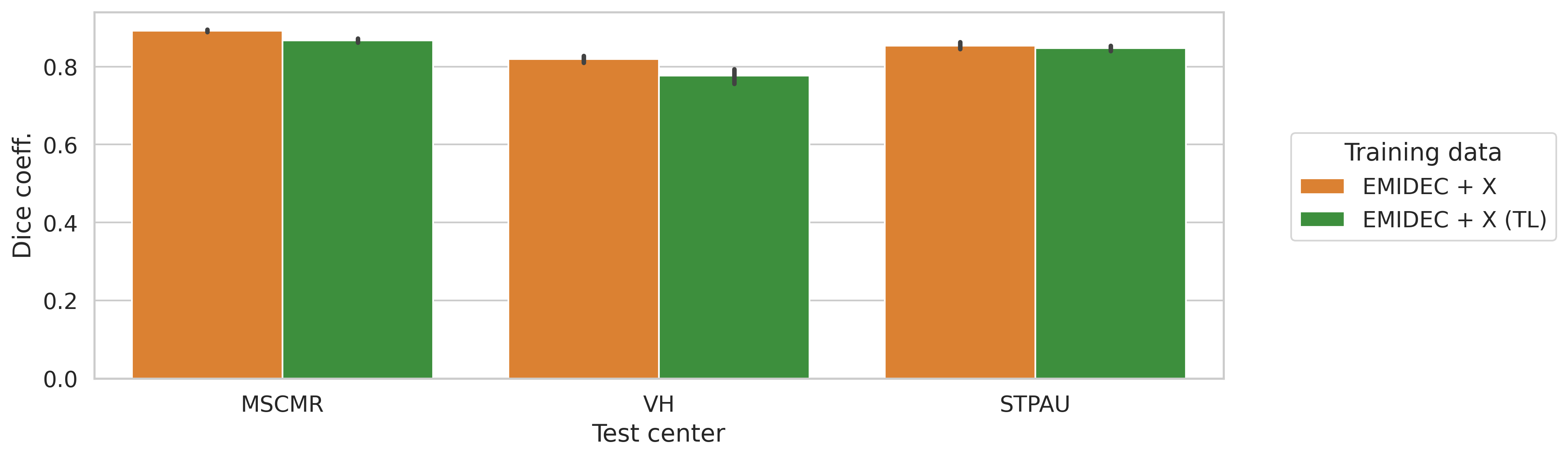}
    \caption{Model trained from scratch using EMIDEC and a second dataset (X),
    which can be MSCMR, VH or STPAU (orange). Then, a model pre-trained with
    EMIDEC, and fine-tuned and evaluated on X (green). The black bars represent
    the standard deviation for five independent runs of each model. TL: transfer
    learning.}
    \label{fig:scratch_vs_tf}
\end{figure}

This section evaluates the potential value of fine-tuning a model pre-trained on
a larger dataset (such as EMIDEC) via transfer learning and the effect of the
sample size used during the tuning process. Figure \ref{fig:blocks} shows the
performance of transfer learning for a model pre-trained with EMIDEC and
fine-tuned for each new clinical center (MSCMR, VH and STPAU). The red and blue
lines in the figure show the segmentation accuracy when the fine-tuning is
performed on the encoder and decoder of the neural network, respectively, while
the remaining parts of the model are frozen. The black line corresponds to a
model trained and tested on the same center. The results show an increase in DC
with the number of fine-tuned blocks and the maximum is obtained when 5 or all
blocks of the encoder are fine-tuned, reaching nearly the same performance as
the single-center model of the new center (black line). Furthermore, in Figure
\ref{fig:scratch_vs_tf}, the single-center models fine-tuned based on 5 encoding
blocks are directly compared to multi-center models trained based on all images
from the original and new clinical centers. Based on the results, fine-tuned
models (green bars) --despite being fine-tuned on the new LGE-MRI images--
achieve similar segmentation performances than models directly trained from
multi-center image data (orange bars). This shows the potential of transfer
learning to adjust and optimize a few layers of the existing single-center model
based on unseen LGE-MRI images from a new clinical center. 

\begin{figure}[hbt!]
    \centering
    \includegraphics[width=.9\textwidth]{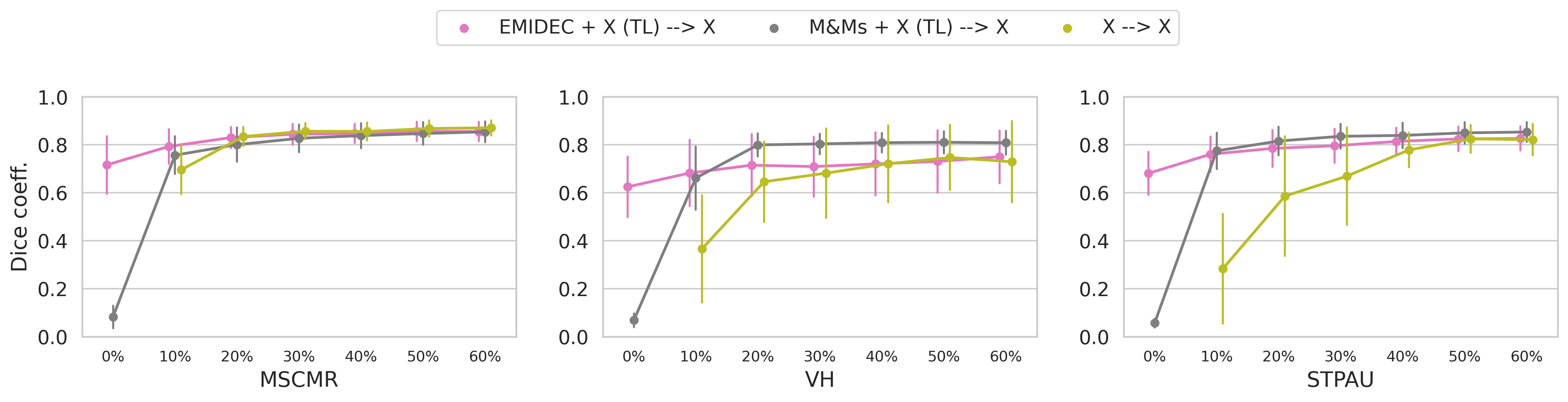}
    \caption{Impact of sample size (percentage) of a new LGE-MRI dataset used
    for fine-tuning existing single-center models for different training
    datasets (fucsia and gray) and compared to single-center models (yellow).
    Results are averaged over five independent model runs.}
    \label{fig:percentages}
\end{figure}

However, transfer learning requires manual annotations of some images from the
new clinical sites. Hence, ideally the number of new annotated images required
to suitably adapt the existing model to the new center should be minimal. In
Figure \ref{fig:percentages}, we evaluated the impact of the number of new
LGE-MRI images used for fine-tuning. The results indicate that the fine-tuning
of single-center models with a small percentage of the target data is sufficient
to reach a desirable segmentation accuracy. Such generalization is achieved, for
example, in the case of a single-center model pre-trained with the EMIDEC
dataset and fine-tuned using only 10\% (about 1-3 subjects) from the new
dataset. A similar pattern is found when training with a different modality, as
shown by the model pre-trained with cine-MRI images from the M\&Ms dataset (gray
line), except for a better performance when testing on the VH center.

\subsection{Experiment 4: Comparison to a multi-center scenario}
\begin{figure}[hbt!]
    \centering
    \includegraphics[width=.9\textwidth]{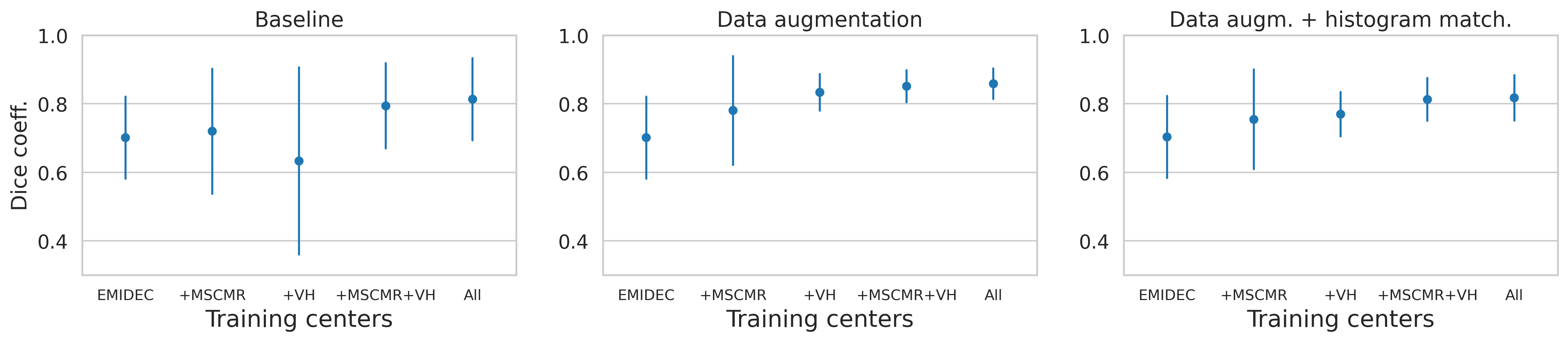}
    \caption{Average DC achieved by models trained on different combinations of
    clinical datasets, with and without data augmentation, as well as with
    histogram matching. The first model is initially trained with the EMIDEC
    dataset, and then new datasets are included progressively from the three
    other clinical sites. Results are averaged over five independent model
    runs.}
    \label{fig:Domain mixing}
\end{figure}

In this last experiment, the added value of training multi-center models for
LGE-MRI segmentation is evaluated by including training images from multiple
clinical sites (i.e. from 1 to 4 centers). In Figure \ref{fig:Domain mixing}
different combinations of the four datasets considered in this study were
explored, either by using a baseline model, data augmentation or histogram
matching. As observed in the results, when the model is trained with no data
augmentation (baseline), the multi-center data enhanced the generalization
ability as demonstrated by the increase in average DC values and the reduction
of the standard deviation. However, when data augmentation is included in the
pipeline, no gain is found by adding new clinical sites to the training stage,
as the data augmentation alone is sufficiently powerful for training the model
with reduced over-fitting when tested in new centers. The results also confirm
that histogram matching does not show significant positive impact on the final
performance.

\subsection{Qualitative analysis}
\begin{figure}[hbt!]
    \centering
    \includegraphics[width=.9\textwidth]{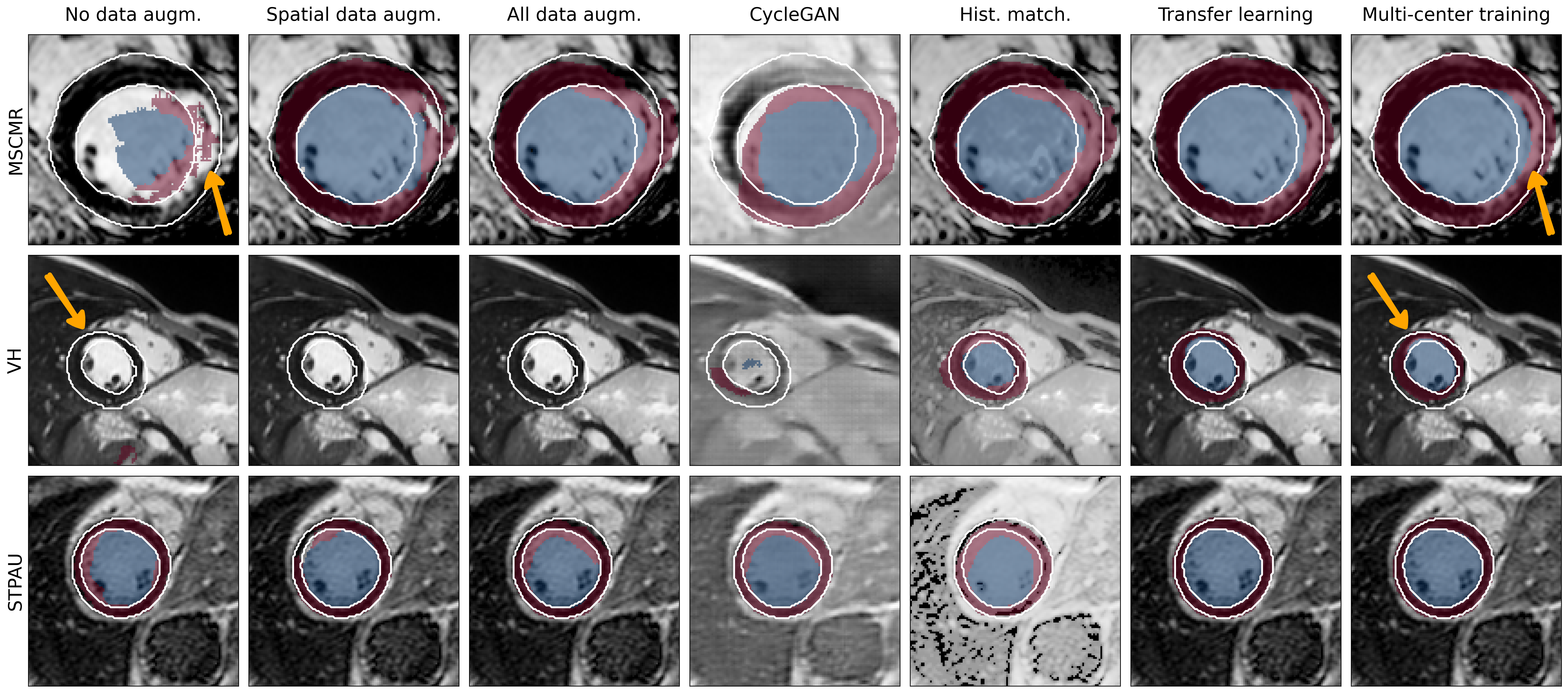}
    \caption{Qualitative comparison of model predictions for selected slices of
    test subjects. The groundtruth is delineated with white lines while the
    overlayed color represents the model prediction. Each row corresponds to a
    different dataset and each column corresponds to each model as presented in
    Table~\ref{tab:results_summary}. Challenges regions with scars are
    highlighted with orange arrows in the first and last columns.}
    \label{fig:qualitative}
\end{figure}

Finally, we show a qualitative comparison in Figure~\ref{fig:qualitative} of
model predictions (colored overlay) for selected cases that demonstrate the
common mistakes of the models as compared to the groundtruth (white
delineations). For instance, the first three columns show how data augmentation
improves the model ability to identify and segment the left ventricle while for
some cases (like for the second row, with the VH sample), it is still
insuficient. The fourth and fifth columns show the effect of the image
harmonization experiments, which help in segmenting failing cases but do not
improve significantly the accuracy of the segmentation as observed in the
disagreement between groundtruth and predictions. Finally, the last two columns
show the predictions for transfer learning and multi-center models,
respectively. These final predictions are the most accurate among all the
columns, but one can still identify some disagreements in challenging regions
annotated with orange arrows were scars can be found.

\section{Discussion}

In this work several strategies were implemented and evaluated for generalizable
segmentation of left ventricular anatomy in multi-center LGE-MRI. The pipeline
was built with the purpose to train single-center models that can maintain a
good level of performance when used to segment out-of-sample images from new
hospitals. The results highlight the importance of using data augmentation,
including both spatial and intensity-based transformations, in particular when
there is a high domain shift between the training and unseen clinical site, e.g.
EMIDEC in our results. After applying adequate data augmentation to existing
single-center models, it was found that neither multi-center training nor image
harmonization techniques are needed to obtain additional generalizability,
confirming the results obtained by \cite{campello2021multi} in the M\&Ms study
for a multi-center and multi-vendor cine-MRI. This finding shows that
single-center LGE-MRI models can generalize well if appropriately enriched with
data augmentation, which results in an important practical benefit: Multi-center
training is difficult in practice as there is a lack of labelling harmonization
between centers, in addition to the legal and other obstacles that make
difficult cross-site data sharing. Moreover, multi-center models are still
specific to those clinical centers that contributed data, whereas there is need
for models that can generalize well beyond the training data. 

Regarding domain adaptation, which theoretically is a promising solution,
existing research has shown that histogram matching could lead to hidden noise
in some images after the post-processing \cite{garg2017comparative}, while
CycleGANs would typically require substantial training data from the new
clinical center to achieve a good model performance. In addition to data
augmentation, the results demonstrated that transfer learning can positively
impact the model performance across sites. This method is based on the
fine-tuning of an existing model initially pre-trained on a single-center
dataset and adjusted  with few datasets from the new clinical site. The obtained
results indicate that fine-tuning the first 5 blocks of the encoder of the model
with the 10\% of the dataset, ranging from 1 to 3 subjects, is sufficient to
achieve the desired LV segmentation accuracy in LGE-MRI. For example, a neural
network pre-trained based on the EMIDEC dataset and fine-tuned with one
subject/image only from STPAU (DC: $0.76\pm 0.07$) performs similarly when
compared to a model trained from scratch with the 100\% of the STPAU images (DC:
$0.79\pm 0.13$). In terms of computational time, the first model is completely
trained in half an hour and the posterior fine-tuning requires only 5 minutes. 

In addition to transfer learning focused on LGE-MRI, we evaluated the potential
of fine-tuning a pre-existing model trained on larger cine-MRI datasets from the
M\&Ms dataset, which consists of a 350 training images. Despite the different
imaging characteristics between cine and LGE-MRI images, in particular the
additional presence of scar tissue and contrast enhancements in the LGE-MRI
images, the results showed that such cross-modality transfer learning results in
enhanced generalizability. This can be easily explained by the fact that such
pre-trained multi-center and multi-disease model encodes additional
inter-subject variability which aids generalizability also in multi-center and
multi-disease LGE-MRI context.

\begin{figure}[hbt!]
\centering
\includegraphics[width=10.5cm]{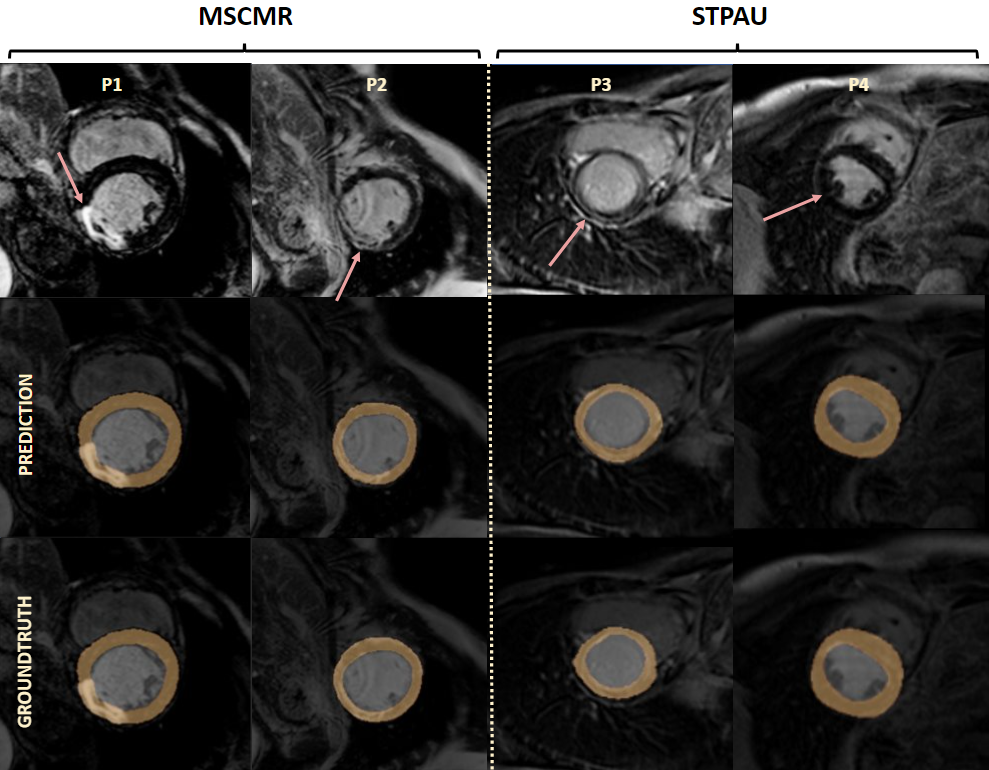}
\caption{\label{fig:path_art_good} Challenging cases leading to good model
predictions on two patients from two different hospitals. First row: original
image, second row: prediction, third row: groundtruth. Each of the two columns
correspond to images obtained from MSCMR or STPAU datasets respectively. The red
arrows highlight the infarct or scar tissue.}
\end{figure}

Finally, to illustrate the success of data augmentation and transfer learning to
build models with good generalization ability, Figure \ref{fig:path_art_good}
provides two examples of challenging LGE-MRI cases, with varying imaging and
anatomical characteristics. Despite the fact that these images are from two
different clinical centers and vary greatly in the appearance, size, shape and
location of the scar tissues, the proposed enriched models are capable to
accurately identify the LV boundaries consistently across the LGE-MRI examples.

Compared to other multi-center existing studies, such as the M\&Ms challenge
that comprises 350 cine-MRI cases, the present multi-center LGE-MRI study has a
lower sample size. This is because the LGE-MRI datasets are less abundant and
more difficult to compile for research studies. Nevertheless, the results in
this work are generated based on 216 datasets from four clinical centers, three
vendors (Siemens, Philips and GE) and three countries from two different
continents. 

Another limitation is that this work was focused on the segmentation of the LV
anatomy and did not consider the more challenging task of segmenting the scar
tissues. This is due to the fact that the clinical annotations for the scar
tissues were not available for the two clinical centers in Spain. Future
multi-center studies in LGE-MRI should also investigate generalizability of
neural networks for scar tissue segmentation. However, our work is an important
first step in this direction, and one that will encourage the development of
more generalizable models based on data augmentation and transfer learning, in
LGE-MRI but also in other cardiac and non-cardiac imaging modalities. 

\begin{figure}[hbt!]
    \centering
    \includegraphics[width=10.5cm]{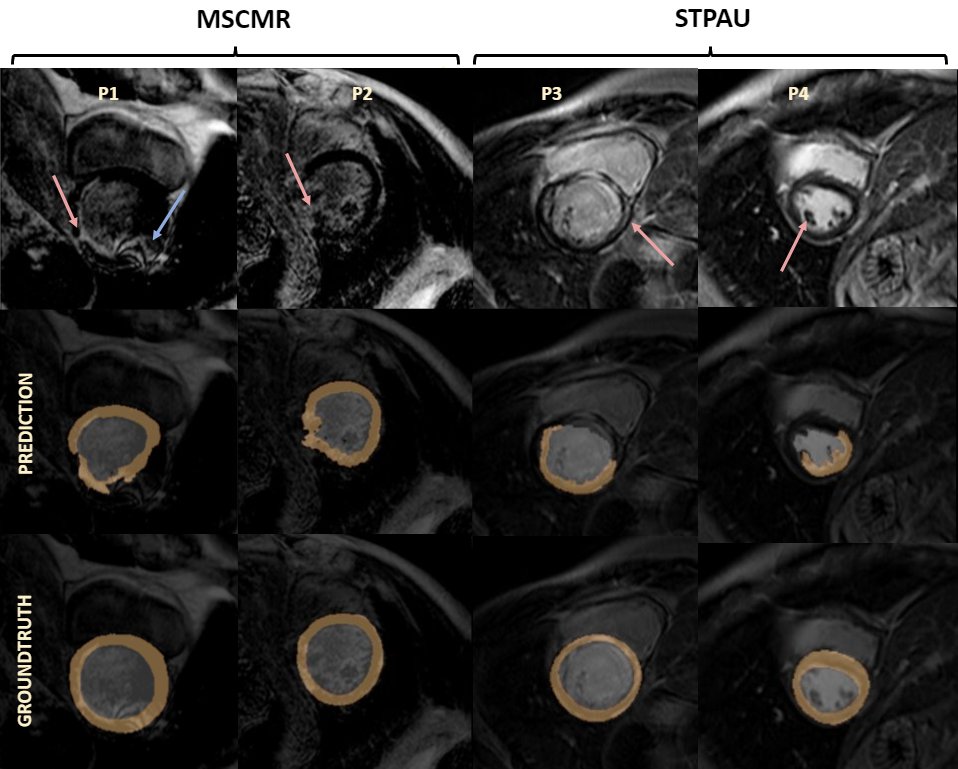}
    \caption{\label{fig:path_art_bad} Segmentation failures obtained due to
    artifacts and highly complex scars. First row: original image, second row:
    prediction, third row: groundtruth. Each of the two columns correspond to
    images obtained from MSCMR and STPAU datasets respectively. The blue arrow
    shows an image artifact, while the red arrows points to the infarct or scar
    tissue.}
    \end{figure}

\begin{figure}[hbt!]
    \centering
    \includegraphics[width=10.5cm]{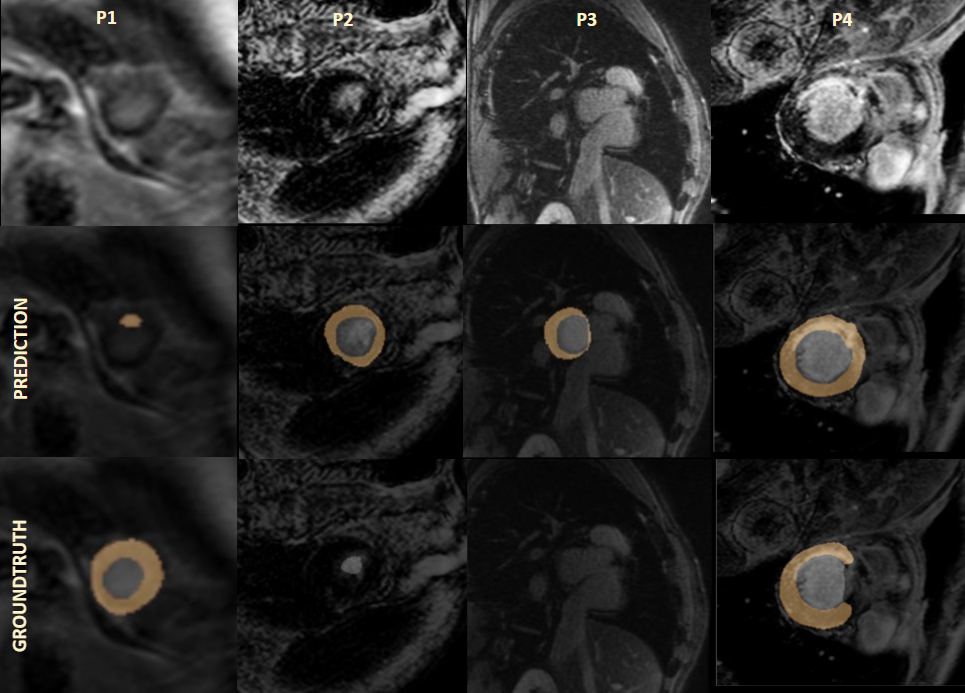}
    \caption{\label{fig:basal_apical}Examples of segmentation failures obtained
    at the apical and basal slices. First row: original image, second row:
    prediction, third row: groundtruth. First and second column show two similar
    cases where both apical slices are segmented differently. The third and
    forth columns are two heterogeneous segmentation at the basal region.}
    \end{figure}

While the proposed framework shows promise for generalizability across
multi-center LGE-MRI datasets with challenging and heterogeneous conditions, it
can fail to accurately identify the LV boundaries in a few exceptions. As
illustrated in Figure \ref{fig:path_art_bad}, a number of failures have been
observed in the presence of low-quality images with artifacts due to suboptimal
contrast wash-out or highly complex scar appearance. Furthermore as reported in
previous works in cardiac cine-MRI segmentation \cite{chen2020improving}, apical
and basal slices are also more error-prone than mid-ventricle slices in LGE-MRI
segmentation. In fact, even experienced cardiologists can disagree on the
segmentation of the LV borders closer to the apex and base, which generates
inter-operator variability that can confuse neural networks, as illustrated in
Figure \ref{fig:basal_apical}.

\section{Conclusions}

This work was motivated by the need for new deep learning based solutions that
generalize well across domains, centers and scans, in non-contrast as well as in
contrast-enhanced imaging. Data augmentation extended the image distribution in
single-center settings and proved to be an effective technique to generate
models with a prominent generalization ability to new clinical centers. In
contrast, image harmonization did not improve the capability of single-center
models when tested on unseen clinical sites. Furthermore, the exploitation of
transfer learning based on fine-tuning of pre-trained models with as little as
one additional subject from an unseen clinical site translated into a
substantial improvement in the model's generalizability. This paper showed that
single-domain neural networks enriched with suitable generalization procedures
can reach  and  even  surpass  the  performance  of  multi-center,  multi-vendor
models in contrast-enhanced imaging,  hence  eliminating the  need  for
comprehensive  multi-center  datasets  to  train  generalizable models. \\
\\
\\
\\
\\
\\
\\
\\
\\
\\
\\
\\
\textbf{Funding}
\\
This work received funding from the European Union's 2020 research and
innovation programme under grant agreement No. 825903 (euCanSHare project), as
well as from the Spanish Ministry of Science, Innovation and Universities under
grant agreement RTI2018-099898-B-I00. Guala A. received funding from the Spanish
Ministry of Science, Innovation and Universities (IJC2018-037349-I). \\
\\
\textbf{Abbreviations}
\\
Late gadolinium-enhanced magnetic resonance imaging (LGE-MRI); Left ventricle
(LV); Magnetic resonance imaging (MRI). \\
\\
\textbf{Availability of data and materials}
\\
EMIDEC and MSCMR dataset are publicly available. The VH and STPAU datasets are
available from the corresponding author upon reasonable request. \\
\\
\textbf{Competing interests}
\\
The authors declare that they have no competing interests. \\
\\
\textbf{Authors' contributions}
\\
Design of the work: CSB, VMC, CMI, KL. Image collection: DVM, MLD, AG, JFRP.
Interpretation of results: CSB, VMC, CMI, KL. Manuscript draft: CSB, KL.
Mansuscript review: VMC, CMI, AG, KL.

\bibliography{elsarticle-template}

\begin{thebibliography}{10}
\expandafter\ifx\csname url\endcsname\relax
  \def\url#1{\texttt{#1}}\fi
\expandafter\ifx\csname urlprefix\endcsname\relax\def\urlprefix{URL }\fi
\expandafter\ifx\csname href\endcsname\relax
  \def\href#1#2{#2} \def\path#1{#1}\fi

\bibitem{guan2021domain}
H.~Guan, M.~Liu, Domain adaptation for medical image analysis: a survey, IEEE
  Transactions on Biomedical Engineering.

\bibitem{chen2020deep}
C.~Chen, C.~Qin, H.~Qiu, G.~Tarroni, J.~Duan, W.~Bai, D.~Rueckert, Deep
  learning for cardiac image segmentation: A review, Frontiers in
  Cardiovascular Medicine 7 (2020) 25.

\bibitem{campello2021multi}
V.~M. Campello, P.~Gkontra, C.~Izquierdo, C.~Mart{\'\i}n-Isla, A.~Sojoudi,
  P.~M. Full, K.~Maier-Hein, Y.~Zhang, Z.~He, J.~Ma, et~al., Multi-centre,
  multi-vendor and multi-disease cardiac segmentation: The m\&ms challenge,
  IEEE Transactions on Medical Imaging.

\bibitem{chen2020improving}
C.~Chen, W.~Bai, R.~H. Davies, A.~N. Bhuva, C.~H. Manisty, J.~B. Augusto, J.~C.
  Moon, N.~Aung, A.~M. Lee, M.~M. Sanghvi, et~al., Improving the
  generalizability of convolutional neural network-based segmentation on cmr
  images, Frontiers in cardiovascular medicine 7 (2020) 105.

\bibitem{Kong2020AGD}
F.~Kong, S.~C. Shadden, A generalizable deep-learning approach for cardiac
  magnetic resonance image segmentation using image augmentation and attention
  u-net, in: Statistical Atlases and Computational Models of the Heart. M{\&}Ms
  and EMIDEC Challenges, Springer International Publishing, Cham, 2021, pp.
  287--296.

\bibitem{Parreno2021}
M.~Parre{\~{n}}o, R.~Paredes, A.~Albiol, Deidentifying mri data domain by
  iterative backpropagation, in: Statistical Atlases and Computational Models
  of the Heart. M{\&}Ms and EMIDEC Challenges, Springer International
  Publishing, Cham, 2021, pp. 277--286.

\bibitem{Corral2021}
J.~Corral~Acero, V.~Sundaresan, N.~Dinsdale, V.~Grau, M.~Jenkinson, A 2-step
  deep learning method with domain adaptation for multi-centre, multi-vendor
  and multi-disease cardiac magnetic resonance segmentation, in: Statistical
  Atlases and Computational Models of the Heart. M{\&}Ms and EMIDEC Challenges,
  Springer International Publishing, Cham, 2021, pp. 196--207.

\bibitem{scannell2020domain}
C.~M. Scannell, A.~Chiribiri, M.~Veta, Domain-adversarial learning for
  multi-centre, multi-vendor, and multi-disease cardiac mr image segmentation,
  in: International Workshop on Statistical Atlases and Computational Models of
  the Heart, Springer, 2020, pp. 228--237.

\bibitem{cheplygina2017transfer}
V.~Cheplygina, I.~P. Pena, J.~H. Pedersen, D.~A. Lynch, L.~S{\o}rensen,
  M.~De~Bruijne, Transfer learning for multicenter classification of chronic
  obstructive pulmonary disease, IEEE journal of biomedical and health
  informatics 22~(5) (2017) 1486--1496.

\bibitem{kushibar2019supervised}
K.~Kushibar, S.~Valverde, S.~Gonz{\'a}lez-Vill{\`a}, J.~Bernal, M.~Cabezas,
  A.~Oliver, X.~Llad{\'o}, Supervised domain adaptation for automatic
  sub-cortical brain structure segmentation with minimal user interaction,
  Scientific reports 9~(1) (2019) 1--15.

\bibitem{Liu2021SemisupervisedMW}
X.~Liu, S.~Thermos, A.~Q. O'Neil, S.~A. Tsaftaris, Semi-supervised
  meta-learning with disentanglement for domain-generalised medical image
  segmentation, in: MICCAI, 2021.

\bibitem{Li2022DomainGO}
C.~Li, Q.~Qi, X.~Ding, Y.~Huang, D.~Liang, Y.~Yu, Domain generalization on
  medical imaging classification using episodic training with task
  augmentation, Computers in biology and medicine 141 (2022) 105144.

\bibitem{carr1984gadolinium}
D.~Carr, J.~Brown, G.~Bydder, R.~Steiner, H.~Weinmann, U.~Speck, A.~Hall,
  I.~Young, Gadolinium-dtpa as a contrast agent in mri: initial clinical
  experience in 20 patients, American Journal of Roentgenology 143~(2) (1984)
  215--224.

\bibitem{PEPE2020101773}
A.~Pepe, J.~Li, M.~Rolf-Pissarczyk, C.~Gsaxner, X.~Chen, G.~A. Holzapfel,
  J.~Egger, Detection, segmentation, simulation and visualization of aortic
  dissections: A review, Medical Image Analysis 65 (2020) 101773.
\newblock \href {http://dx.doi.org/https://doi.org/10.1016/j.media.2020.101773}
  {\path{doi:https://doi.org/10.1016/j.media.2020.101773}}.

\bibitem{otto2012practice}
C.~Otto, The Practice of Clinical Echocardiography, ClinicalKey 2012,
  Elsevier/Saunders, 2012.

\bibitem{riederer2018technical}
S.~J. Riederer, E.~G. Stinson, P.~T. Weavers, Technical aspects of
  contrast-enhanced mr angiography: current status and new applications,
  Magnetic Resonance in Medical Sciences 17~(1) (2018) 3.

\bibitem{ferre2012advanced}
J.-C. Ferr{\'e}, M.~S. Shiroishi, M.~Law, Advanced techniques using contrast
  media in neuroimaging, Magnetic Resonance Imaging Clinics 20~(4) (2012)
  699--713.

\bibitem{onishi2020ultrafast}
N.~Onishi, M.~Sadinski, M.~C. Hughes, E.~S. Ko, P.~Gibbs, K.~M. Gallagher,
  M.~M. Fung, T.~J. Hunt, D.~F. Martinez, A.~Shukla-Dave, et~al., Ultrafast
  dynamic contrast-enhanced breast mri may generate prognostic imaging markers
  of breast cancer, Breast Cancer Research 22~(1) (2020) 1--13.

\bibitem{welle2020mri}
C.~L. Welle, F.~F. Guglielmo, S.~K. Venkatesh, Mri of the liver: choosing the
  right contrast agent, Abdominal radiology 45~(2) (2020) 384--392.

\bibitem{uhlig2020gadolinium}
J.~Uhlig, O.~Al-Bourini, R.~Salgado, M.~Francone, R.~Vliegenthart,
  J.~Bremerich, J.~Lotz, M.~Gutberlet, Gadolinium-based contrast agents for
  cardiac mri: use of linear and macrocyclic agents with associated safety
  profile from 154 779 european patients, Radiology: Cardiothoracic Imaging
  2~(5) (2020) e200102.

\bibitem{doltra2013emerging}
A.~Doltra, B.~Hoyem~Amundsen, R.~Gebker, E.~Fleck, S.~Kelle, Emerging concepts
  for myocardial late gadolinium enhancement mri, Current cardiology reviews
  9~(3) (2013) 185--190.

\bibitem{yue2019cardiac}
Q.~Yue, X.~Luo, Q.~Ye, L.~Xu, X.~Zhuang, Cardiac segmentation from lge mri
  using deep neural network incorporating shape and spatial priors, in:
  International Conference on Medical Image Computing and Computer-Assisted
  Intervention, Springer, 2019, pp. 559--567.

\bibitem{zabihollahy2020fully}
F.~Zabihollahy, M.~Rajchl, J.~A. White, E.~Ukwatta, Fully automated
  segmentation of left ventricular scar from 3d late gadolinium enhancement
  magnetic resonance imaging using a cascaded multi-planar u-net (cmpu-net),
  Medical physics 47~(4) (2020) 1645--1655.

\bibitem{kurzendorfer2019left}
T.~Kurzendorfer, K.~Breininger, S.~Steidl, A.~Maier, R.~Fahrig, Left ventricle
  segmentation in lge-mri using multiclass learning, in: Medical Imaging 2019:
  Image Processing, Vol. 10949, International Society for Optics and Photonics,
  2019, p. 1094929.

\bibitem{zhuang2020cardiac}
X.~Zhuang, J.~Xu, X.~Luo, C.~Chen, C.~Ouyang, D.~Rueckert, V.~M. Campello,
  K.~Lekadir, S.~Vesal, N.~RaviKumar, et~al., Cardiac segmentation on late
  gadolinium enhancement mri: a benchmark study from multi-sequence cardiac mr
  segmentation challenge, arXiv preprint arXiv:2006.12434.

\bibitem{lalande2020emidec}
A.~Lalande, Z.~Chen, T.~Decourselle, A.~Qayyum, T.~Pommier, L.~Lorgis, E.~de~la
  Rosa, A.~Cochet, Y.~Cottin, D.~Ginhac, et~al., Emidec: A database usable for
  the automatic evaluation of myocardial infarction from delayed-enhancement
  cardiac mri, Data 5~(4) (2020) 89.

\bibitem{zhuang2016multivariate}
X.~Zhuang, Multivariate mixture model for cardiac segmentation from
  multi-sequence mri, in: International Conference on Medical Image Computing
  and Computer-Assisted Intervention, Springer, 2016, pp. 581--588.

\bibitem{zhuang2018multivariate}
X.~Zhuang, Multivariate mixture model for myocardial segmentation combining
  multi-source images, IEEE transactions on pattern analysis and machine
  intelligence 41~(12) (2018) 2933--2946.

\bibitem{zhu2017unpaired}
J.-Y. Zhu, T.~Park, P.~Isola, A.~A. Efros, Unpaired image-to-image translation
  using cycle-consistent adversarial networks, in: Proceedings of the IEEE
  international conference on computer vision, 2017, pp. 2223--2232.

\bibitem{ma2019neural}
C.~Ma, Z.~Ji, M.~Gao, Neural style transfer improves 3d cardiovascular mr image
  segmentation on inconsistent data, in: International Conference on Medical
  Image Computing and Computer-Assisted Intervention, Springer, 2019, pp.
  128--136.

\bibitem{isensee2021nnu}
F.~Isensee, P.~F. Jaeger, S.~A. Kohl, J.~Petersen, K.~H. Maier-Hein, nnu-net: a
  self-configuring method for deep learning-based biomedical image
  segmentation, Nature Methods 18~(2) (2021) 203--211.

\bibitem{garg2017comparative}
P.~Garg, T.~Jain, A comparative study on histogram equalization and cumulative
  histogram equalization, International Journal of New Technology and Research
  3~(9) (2017) 263242.

\end{thebibliography}

\end{document}